\documentclass[11pt,draft]{article}

\usepackage{amsmath}

\begin{document}

\renewcommand{\thefootnote}{\fnsymbol{footnote}}

\begin{center}
\textbf{\large On the thermodynamical limit of self-gravitating systems } \\
\vspace{0.5truecm}
Victor Laliena\footnote{E-mail: laliena@posta.unizar.es \\ 
\hspace*{0.5truecm} Phone: +34 976 761257 \\ 
\hspace*{0.5truecm} Fax: +34 976 761264} \\
\vspace{0.25truecm}
\textit{Departamento de F\'{\i}sica Te\'orica, Universidad de
Zaragoza, \\
Cl. Pedro Cerbuna 12, E-50009 Zaragoza (Spain)} \\
\vspace{0.25truecm}
March 11, 2003 \\
\vspace{0.5truecm}
astro-ph/0303301
\end{center}

\vspace{1truecm}

\begin{center}
ABSTRACT
\end{center}

It is shown that the diluted thermodynamical limit of a self-gravitating
system proposed by de~Vega and S\'anchez suffers from the same problems as
the usual thermodynamical limit and leads to divergent thermodynamical
functions. This question is also discussed from the point of view of mean 
field theory.

\vspace{1truecm}

\vfill\eject

\renewcommand{\thefootnote}{\arabic{footnote}}
\setcounter{footnote}{0}

Recently, de Vega and S\'anchez have pointed out that a kind of
thermodynamical limit of a self-gravitating
system can be defined if one considers what they call the
\textit{diluted} limit: send the number of particles, $N$, and
the volume, $V$, to infinity, keeping constant the ratio $N/V^{1/3}$
instead of the density, $N/V$ \cite{devega}.
This is a rather surprising --and very interesting-- suggestion, since
it is well known that the usual thermodynamical limit of self-gravitating
systems leads to singular thermodynamical functions (due to the gravitational 
instability) \cite{pad,kiess}. However,
we will show in this paper that the diluted regime leads to divergent
thermodynamical functions in a way similar to the usual thermodynamical
limit. Before demonstrating this statement, we will remember the 
differences, in what concerns the thermodynamical limit,
between a thermodynamical stable system \cite{stability} 
and a self-gravitating system, 
and present the arguments that support the idea that the diluted limit 
gives well defined thermodynamical functions for self-gravitating systems. 
Afterwards, we shall proof that this is not possible, and we will
point out the loophole in the arguments that lead to the wrong conclusion.
We will end the paper with a brief discussion on mean field theory.
\\

Consider a system of $N$ classical particles, endowed with a hard core,
the interactions of which are described
by a potential energy which is a sum of two body
contributions:
\begin{equation}
\Phi_N(r_1,\ldots,r_N)=\sum_{i<j}\phi(r_i-r_j) \, .
\end{equation}
Using the notation $\phi_{ij}\equiv\phi(r_i-r_j)$, 
the microcanonical and canonical partition functions, $\mathcal{Z}_{MC}$
and $\mathcal{Z}_{C}$, respectively, can be written as
\begin{eqnarray}
\mathcal{Z}_\mathrm{MC} &=& \frac{1}{N! \Gamma(3N/2)}\,\int_{V^N}\,
\prod_{i=1}^N d^3r_i\,[E-\sum_{i<j}\phi_{ij}]_+^{3N/2-1}\,
\, , \label{mcpf} \\
\mathcal{Z}_\mathrm{C} &=& \frac{1}{N!}\,\beta^{-3N/2}\,\int_{V^N}\,
\prod_{i=1}^N d^3r_i\,\exp(-\beta \sum_{i<j}\phi_{ij}) \label{cpf} \, , 
\end{eqnarray}
where $E$ is the energy, $V$ the volume, $\beta$ the inverse temperature,
$[x]_+=x$ if $x>0$ and
$[x]_+=0$ if $x\leq 0$, and we have ignored some
irrelevant factors involving powers of the particle mass.
Introducing the family of functions
\begin{equation}
G^{(\alpha)}_V(u)= \int_{V^N}\,\prod_{i=1}^N d^3r_i\,
\delta (u-\frac{1}{N^\alpha}\sum_{i<j}\phi_{ij}) \, ,
\end{equation}
the partition functions can be obviously written as
\begin{eqnarray}
\mathcal{Z}_\mathrm{MC} &=& \frac{1}{N! \Gamma(3N/2)}\,
\int\,du\,G^{(\alpha)}_V(u)\,[E-N^\alpha u]_+^{3N/2-1}\,
\, , \\
\mathcal{Z}_\mathrm{C} &=& \frac{1}{N!}\,\beta^{-3N/2}\,
\int\,du\,G^{(\alpha)}_V(u) \exp[-\beta N^\alpha u] \, .
\end{eqnarray}

Assuming that, as $N\rightarrow\infty$, keeping the density $\rho=N/V$ fixed,
the following asymptotic behavior holds 
\begin{equation}
G^{(1)}_V(u) \approx e^{N g(u,\rho)} \, ,
\label{scaling:short}
\end{equation}
we can get the partition functions with the aid of the saddle point method.
For the microcanonical ensemble (MC) we have the equation
\begin{equation}
\frac{\partial g(u,\rho)}{\partial u} = \frac{3}{2}\frac{1}{E/N-u} \, .
\end{equation}
The solution is, obviously, a function of $\epsilon=E/N$ and $\rho=N/V$, 
$u=\bar{u}_m(\epsilon,\rho)$, and the entropy scales with $N$.
The corresponding saddle point equation for the canonical ensemble (CE) 
is 
\begin{equation}
\frac{\partial g(u,\rho)}{\partial u} = \beta \, ,
\end{equation}
the solution of which is a function of $\beta$ and $\rho$,
$u=\bar{u}_c(\beta,\rho)$, and the corresponding thermodynamical
potential (usually called the Helmholtz free energy) is extensive.
It is not difficult to realize that the saddle point equations
imply the equivalence between the MC and the CE.
\\

Let us formally see that the scaling (\ref{scaling:short}) holds
if the two body potential is short ranged. We will use the notation 
\begin{equation}
\langle f(r_1,\ldots,r_N)\rangle = \frac{1}{V^N}\,\int_{V^N}\,
\prod_i d^3r_i\,f(r_1,\ldots,r_N) \, . 
\label{definition}
\end{equation}
Using the Fourier representation of the Dirac delta we can write
\begin{equation}
G^{(\alpha)}_V(u)=V^N \int_{-\infty}^\infty\frac{d\omega}{2\pi}\,
e^{\mathrm{i} \omega u}\,\tilde{G}_V^{(\alpha)}(\omega)
\label{fourier}
\end{equation}
where, with our definition (\ref{definition}),
\begin{equation}
\tilde{G}_V^{(\alpha)}(\omega)\equiv
\langle\,\exp(-\mathrm{i}\frac{\omega}{N^\alpha}\sum_{i<j}\phi_{ij})\,
\rangle = \frac{1}{V^N}
\int_{V^N}\,\prod_{i=1}^N\,d^3r_i\,
\exp(-\mathrm{i} \frac{\omega}{N^\alpha} \sum_{i<j}\phi_{ij}) \, .
\end{equation}
We can now apply the cumulant expansion to the above expression 
\begin{equation}
\langle\,\exp(-\mathrm{i}\omega\sum_{i<j}\phi_{ij})\,\rangle = 
\exp\left(\sum_{n=0}^\infty 
\frac{(-\mathrm{i}\omega)^n}{n! N^{\alpha n}}  
\sum_{i_1<j_1}\ldots\sum_{i_n<j_n}
\langle\phi_{i_1 j_1}\cdots\phi_{i_n j_n}\rangle_\mathrm{c}\right)\, ,
\end{equation}
where
$\langle\phi_{i_1 j_1}\cdots\phi_{i_n j_n}\rangle_\mathrm{c}$ is the 
connected correlation function.
It is well known that 
only those sequences of coordinates
$\{i_1j_1,\ldots,i_nj_n\}$ for which the integral of Eq.~(\ref{intpot})
below cannot be split into the product of two (or more) integrals
contribute to the connected correlation function. We will call
such sequences connected sequences. 
For large $N$ and $n\ll N$, the number of connected sequences 
asymptotically is
$c_n N^{n+1}$, where $c_n$ is a number independent of $N$.
Let us define $\xi_l=r_{i_l}-r_{j_l}$, for $l=1,\ldots,n$.
The number of connected sequences for which the $\xi_l$'s are
linearly dependent scales with a lower power of $N$, and, therefore,
they do not
contribute as $N\rightarrow\infty$. Only connected sequences with
the $\xi_l$'s linearly independent contribute to the leading term
in $N$ of the $n$-th order of the cumulant expansion. For such
sequences, if the potential decays sufficiently fast at long distances, the
integral of $\phi_{i_1j_1}\cdots\phi_{i_nj_n}$ over the $\xi_l$'s
gives a number, $\chi_{i_1j_1\ldots i_nj_n}$, that is independent
of the volume if $V$ is large. The integral over the remaining variables
gives a power of the volume, $V^{N-n}$. Therefore, we have:
\begin{equation}
\frac{1}{V^N}\,
\int_{V^N}\,\prod_{i=1}^N\,d^3r_i\, \phi_{i_1j_1}\cdots\phi_{i_nj_n}
\approx \frac{\chi_{i_1j_1\ldots i_nj_n}}{V^n} \, .
\label{intpot}
\end{equation}
The connected correlation function is a sum of products of
integrals of the above type and has the same behavior.
Collecting all the results, we have
\begin{equation}
\sum_{i_1<j_1}\ldots\sum_{i_n<j_n}
\langle\phi_{i_1 j_1}\cdots\phi_{i_n j_n}\rangle_\mathrm{c} \approx
N^{n+1}\,\frac{c_n \chi_n}{V^n} \, ,
\end{equation}
where $\chi_n$ is the average of 
$\chi_{i_1j_1 \ldots i_nj_n}$
over the connected sequences.
This behavior is broken down for $n\sim N$, but the contribution of these
terms to the cumulant expansion is negligible.
Then, the $n$-th term is
\begin{equation}
(-\mathrm{i})^n \frac{c_n \chi_n}{n!} 
\frac{\omega^n}{N^{\alpha n-1}}\left(\frac{N}{V}\right)^n \, .
\end{equation}
Performing the change of variables 
$\omega\rightarrow N \omega$ in
the integral of Eq.~(\ref{fourier}), we get for $\alpha=1$ 
\begin{equation}
G^{(1)}_V(u)\approx \frac{V^N}{N}\int_{-\infty}^\infty\frac{d\omega}{2\pi}\,
e^{N[\mathrm{i} \omega u + \Omega(\mathrm{i} \omega \rho)]}\, ,
\label{fourierass}
\end{equation}
where 
\begin{equation}
\Omega(x)=\sum_{n=0}^\infty (-1)^n\frac{c_n \chi_n}{n!} x^n \, .
\label{series}
\end{equation}
The integral of Eq.~(\ref{fourierass}) can be evaluated by the saddle
point method, and gives
\begin{equation}
G^{(1)}_V(u)\approx\frac{V^N}{N}\exp[N g(u/\rho)] \, ,
\end{equation}
where
\begin{equation}
g(u/\rho)=\bar{\omega}(u/\rho) u/\rho + \Omega[\bar{\omega}(u/\rho)]\, ,
\end{equation}
and $\bar{\omega}(u/\rho)$ is the solution of
\begin{equation}
\Omega'(\bar{\omega})=-u/\rho \, .
\end{equation}
Therefore, if the potential is short ranged, $G^{(1)}(u)$ verifies the
scaling of Eq.~(\ref{scaling:short}) and the thermodynamical functions
are well behaved in the thermodynamical limit.
\\

What about a self-gravitating system? Since the potential decays as
the inverse distance, we have that, in the same conditions as before
\begin{equation}
\frac{1}{V^N}\int_{V^N}\,\prod_{k=1}^N\,d^3r_k\,
\phi_{i_1j_1}\cdots\phi_{i_nj_n}
\approx\frac{\chi_{i_1j_1\ldots i_nj_n}}{R^n} \, ,
\end{equation}
where $R$ is the linear size of the system ($V=R^3$). The same arguments
of the above paragraphs lead to the following expression for the $n$-th  
term of the cumulant expansion:
\begin{equation}
(-\mathrm{i})^n \frac{\chi_n}{n!} 
\frac{\omega^n}{N^{\alpha n-1}} V^{2n/3}\left(\frac{N}{V}\right)^n \, .
\label{nthself}
\end{equation}
The above expression shows that, as expected, the asymptotic 
behavior~(\ref{scaling:short}) does not hold.
However, we see that for $\alpha=5/3$ the $n$-th order term of the
cumulant expansion is
\begin{equation}
(-\mathrm{i})^n \frac{\chi_n}{n!} 
\frac{\omega^n}{N^{n-1}} \left(\frac{N}{V}\right)^{n/3} \, .
\end{equation}
Hence, changing the variable $\omega\rightarrow N \omega$ in
the integral of Eq.~(\ref{fourier}) and using the saddle point method,
leads to
\begin{equation}
G^{(5/3)}_V(u)\approx\frac{V^N}{N}\exp[N g(u/\rho^{1/3})] 
\end{equation}
for a self-gravitating system.

Now, the saddle point equation that gives the MC partition function is 
\begin{equation}
\frac{\partial g(u/\rho^{1/3})}{\partial u} = 
\frac{3}{2}\frac{1}{E/N^{5/3}-u} \, .
\end{equation}
The solution is a function of $E/N^{5/3}$ and $\rho$, 
$u=\rho^{1/3} \bar{u}_\mathrm{m}[E/(\rho^{1/3} N^{5/3})]$, and therefore
the entropy is not extensive\footnote{It is not an homogeneous function of
$N$, $E$, and $V$.}.
The MC temperature, 
$T_\mathrm{MC}=(\partial \ln \mathcal{Z}_\mathrm{MC}/\partial E)^{-1}$, is 
\begin{equation}
T_\mathrm{MC}=\frac{2E}{3N} - 
\frac{2}{3} N^{2/3} \rho^{1/3}\bar{u}_\mathrm{m}[E/(N^{5/3}\rho^{1/3})]\, .
\end{equation}
We see that the MC temperature diverges as the number of particles
increases. This is a manifestation of the so called gravothermal catastrophe
\cite{gravothermal}.
 
Let us analyze the canonical ensemble.
If the temperature is fixed, the contribution of $G^{(5/3)}_V(u)$ is negligible
compared with the contribution of the Boltzmann weight, 
$\exp[-\beta N^{5/3} u]$, and the saddle point solution is
given by $u=\bar{u}_\mathrm{c}=u_0$, where $u_0$ is the minimum possible value
of $u$, that is, the minimum potential energy. Hence, the system is 
completely collapsed. The only way to avoid the complete collapse of
the system as the number of particles
increases is to increase the temperature by a factor $N^{2/3}$, in agreement
with the MC analysis.

The thermodynamical potentials for self-gravitating systems
are non extensive and the system will be inhomogeneous. The thermodynamical
limit is singular and gives ill behaved thermodynamical functions.
\\

Up to here, everything is known.
There is, however, another interesting possibility:
as de Vega and S\'anchez pointed out \cite{devega}, it seems that the function
$G^{(1)}_V(u)$ will have the asymptotic behavior as $\exp[N g(u,N/R)]$ if
we keep $\sigma=N/R$ constant (instead of $\rho=N/V$) as $N\rightarrow\infty$.
They called this the diluted limit. Indeed, plugging $V=R^3$ in 
Eq.~(\ref{nthself}), the $n$-th term of the
cumulant expansion of $\tilde{G}^{(1)}_V(\omega)$ reads
\begin{equation}
(-\mathrm{i})^n \frac{\chi_n}{n!} 
\frac{\omega^n}{N^{n-1}} \left(\frac{N}{R}\right)^n \, .
\label{nthdil}
\end{equation}
Then, $G^{(1)}_V(u)\approx \exp[N g(u,\sigma)]$, and the thermodynamical
potentials will scale with $N$. In particular, the canonical 
partition function will scale as
\begin{equation}
\mathcal{Z}_\mathrm{C} \approx\frac{V^N}{N} e^{-N f(\beta,\sigma)} \, .
\end{equation}
The system will not be extensive, 
however, and will develop inhomogeneities, but the 
thermodynamical functions will be well behaved in the thermodynamical
(diluted) limit \cite{devega}.
\\

The above proof of the existence of the diluted thermodynamical limit
is formal, since it relies on a series expansion, the convergence of which
has not been demonstrated. Indeed, we are going to
show that the diluted limit cannot give well behaved
thermodynamical functions.
Let us consider the canonical partition function on a volume $V=R^3$
in the diluted regime, so that $N\sim R$, and let us take a portion
of such volume of linear size $R_0<R$, such that $N\sim R_0^3$.
Then, we obviously have
\begin{equation}
\mathcal{Z}_\mathrm{C}  \geq
\frac{1}{N!}\,\beta^{-3N/2}\,\int_{V_0^N}\,
\prod_{i=1}^N d^3r_i\,\exp[-\beta \sum_{i<j}\phi_{ij}] \geq
\frac{V_0^N}{N!} \exp[-\beta \sum_{i<j} \langle \phi_{ij} \rangle_{V_0}] \, ,
\end{equation}
where
\begin{equation} 
\langle\phi_{ij}\rangle_{V_0}=\frac{1}{V_0^N}\,
\int_{V_0^N}\prod_{k=1}^N d^3r_k \, \phi_{ij} \, , 
\end{equation}
and the last inequality follows from a well known property of the exponential 
function:
$\langle \exp(y)\rangle \geq \exp(\langle y\rangle)$.
Since $\langle\phi_{ij}\rangle_{V_0}=-\kappa/R_0$, where $\kappa>0$ is a
geometrical number independent of $R_0$ if $R_0$ is large, 
we have
\begin{equation}
\mathcal{Z}_\mathrm{C} \geq \frac{V_0^N}{N!} e^{\beta N(N-1)\kappa /R_0} \, .
\label{ineq}
\end{equation}
Recalling that $R_0\sim L_0 N^{1/3}$, where $L_0$ is a fixed number with
dimensions of length, we have
\begin{equation}
\mathcal{Z}_\mathrm{C} \geq \frac{N^N}{N!} \exp[N^{5/3} \beta \kappa/L_0] \, .
\label{asympbound}
\end{equation}
Therefore, the canonical partition function cannot behave as 
$\exp[-N f(\beta,\sigma)]$ in the diluted thermodynamical limit. 
Rather, it behaves as in the usual thermodynamical limit.
The reason is clear: the partition function is dominated by collapsed
configurations even in the diluted regime. The gain in entropy provided
by the dilution, which is of the order $N\ln V$, cannot compete with the
energy gain due to collapse, which of the order $N^{5/3}$. 

If the canonical partition function does not scale as 
$\exp[-N f(\beta,\sigma)]$,
it is impossible that $G^{(1)}_V(u)\approx \exp[N g(u,\sigma)]$.
This is in conflict with the result of the cumulant expansion. The 
inequality~(\ref{ineq})
that originates this conflict is rigorous, so that the fallacy must
be found in the cumulant expansion. The solution of the paradox is that the
cumulant expansion for $G^{(1)}_V(u)$
is dominated by terms of the order of $N$. It should converge for any
finite $N$, but the radius of convergence shrinks to zero as 
$N\rightarrow\infty$.
Indeed, $\tilde{G}_V^{(1)}(N \omega)$ is the analytical continuation of
the $\mathcal{Z}_\mathrm{C}(\beta)$ to the imaginary axis. 
If the radius of convergence of the cumulant expansion were finite in
the diluted thermodynamical limit, it would imply [\textit{cf.} 
Eq.~(\ref{fourierass})]
\begin{equation}
\tilde{G}_V^{(1)}(N \omega)\approx e^{N\Omega(\mathrm{i}\omega\sigma)} \, .
\end{equation}
The asymptotic behavior of the above equation is not compatible with
Eq.~(\ref{asympbound}). The cumulant expansion for $\Omega(x)$
given by Eq.~(\ref{series}) must have a vanishing convergence radius
in the diluted thermodynamical limit of a self-gravitating system.

There is a clear explanation of the failure of the cumulant expansion.
For imaginary $\omega$, $\tilde{G}_V^{(1)}(N \omega)$ is a canonical
partition function at temperature $1/\mathrm{Im}(\omega)$.
The cumulant expansion relies on connected correlation functions
computed with a flat measure, $V^{-N}\prod_i d^3r_i$, instead
of the Boltzmann weight, $\mathcal{Z}_\mathrm{C}^{-1}\prod_i d^3r_i
\exp[-\mathrm{Im}(\omega)\sum \phi_{ij}]$. The cumulant expansion, 
then, will be valid when the two measures are similar, \textit{i.e.},
in a gas phase. In a self-gravitating system, Eq.~(\ref{asympbound})
indicates that the canonical partition function is completely dominated
by the potential energy, and therefore that collapse takes place,
at temperatures smaller or of the order of $N^{2/3}$. Hence, 
for imaginary $\omega$, the cumulant expansion will be only valid for 
$\mathrm{Im}(\omega) < N^{-2/3}$. Thus, the convergence radius of
the cumulant expansion shrinks to zero as $N^{-2/3}$ in the diluted
thermodynamical limit. 

The collapse in the thermodynamical limit can be avoided  
by rescaling properly the radius of the hard core, $a$ 
\footnote{I am grateful to P.-H. Chavanis and O. Fliegans for 
pointing this out to me.}.
Indeed, it is known that in mean field theory the collapse phase
covers the whole phase diagram in the limit in which the 
''filling'' parameter, $N a^3/R^3$,  tends to zero \cite{chavanis}. 
The ''filling'' parameter remains constant in the usual
thermodynamical limit. To keep it constant in the diluted limit,
the hard core must be scaled as $a\sim N^{2/3}$. Then, previous argument
does not apply, since such big particles do not fit in a volume of
linear size $R_0\sim N^{1/3}$. The minimum size of a region able to
enclose the $N$ particles must have a linear size scaling as $N$.
Hence, the diluted thermodynamical limit can exist. This is,
however, a rather trivial and \textit{ad hoc} way of avoiding
collapse: particles are forced to remain far away one from another,
and it is physically difficult to justify the scaling of the hard
core. Moreover, this way of preventing collapse in the thermodynamical
limit cannot apply to other types of short distance regularizations,
such as softened potentials \cite{edu}. Our proof of the non-existence
of the diluted thermodynamical limit is also valid for softened potentials.
\\

It is generally believed that mean field theory is exact for 
self-gravitating systems. Since in mean field theory any thermodynamical 
function depends on the thermodynamical variables only through the
dimensionless combinations $\Lambda=-ER/(Gm^2N^2)$ (MC) or
$\eta=\beta Gm^2N/R$ (CE), the thermodynamical
limit must be taken keeping either $\Lambda$ or $\eta$ finite.
Then, it could be argued that the dependence of $\Lambda$ or $\eta$ on
$N$ can be absorbed in any of the quantities entering $\Lambda$ or $\eta$.
For instance, one could take $R\propto N$ and $E\propto N$, as in the
diluted limit. We will show in the following that this is not correct.

Consider a softened
gravitational potential, for instance $-Gm^2\bar\phi(r)/R$, with
$\bar\phi(r)=R/\sqrt{r^2+s^2}$, where $s$ is the softening scale.  
Dividing the finite region that encloses
the system in $W$ cells of linear size $w$ ($w^3=R^3/W$), 
and denoting by $n_i$ the occupation number of the $i$th cell,
the partition functions
(\ref{mcpf})~and~(\ref{cpf}) can be approximated by~\cite{pad}:
\begin{eqnarray}
\mathcal{Z}_\mathrm{MC} &=& \frac{1}{\Gamma(3N/2)}\,
\sum_{n_1=0}^N\ldots\sum_{n_W=0}^N
\,\frac{\delta(\sum_l n_l-N)}{\prod_l n_l!}\,
[-\Lambda+\sum_{i,j} n_i n_j\bar\phi_{ij}]_+^{3N/2} , \hspace{1truecm}
\\
\mathcal{Z}_\mathrm{C} &=& \frac{1}{\Gamma(3N/2)}\,
\sum_{n_1=0}^N\ldots\sum_{n_W=0}^N
\,\frac{\delta(\sum_l n_l-N)}{\prod_l n_l!}\,
\exp[\eta\sum_{i,j} n_i n_j\bar\phi_{ij}]\, ,
\end{eqnarray}
where $\bar\phi_{ij}=\bar\phi(r_i-r_j)$ and $r_i$ is the position of
the center of the $i$th cell (we have again ignored constant factors).
The above equalities hold rigorously in the limit $W\rightarrow\infty$.
For large $N$, the factorials can be approximated by their asymptotic form.
Defining the density by $\rho_i=n_i/(N w^3)$, so that 
$\rho_i\in [0,1/w^3]$ becomes a continuum variable as $N\rightarrow\infty$, 
we have, ignoring constant factors
\begin{eqnarray}
\mathcal{Z}_\mathrm{MC} &=& 
\int\,\prod_{k=1}^W\,d\rho_k\,\delta(\sum_l w^3 \rho_l-1) \times 
\hspace{6.25truecm}\nonumber \\
 && \exp\{N[-w^3\sum_i(\rho_i\ln\rho_i-\rho_i)+
\frac{3}{2}
\ln[-\Lambda+\sum_{i,j} w^6 \rho_i\rho_j\bar\phi_{ij}]_+]\}
\, , \label{dmcpf}
\\
\mathcal{Z}_\mathrm{C} &=& 
\int\,\prod_{k=1}^W\,d\rho_k\,\delta(\sum_l w^3 \rho_l-1) \times \nonumber \\
& & \exp\{N[-w^3\sum_i(\rho_i\ln\rho_i-\rho_i)+
\eta\sum_{i,j}w^6\rho_i \rho_j\bar\phi_{ij}]\}\, .
\label{dcpf}
\end{eqnarray}

Mean field theory is obtained by taking $N\rightarrow\infty$ 
before $W\rightarrow\infty$. In such case, the integrals in 
the above equations are saturated by the maximum of the integrands, 
provided that the number of cells,
$W$,  and $\Lambda$ or $\eta$, respectively, 
are kept constant. Obviously, the maximum of the integrand
is given by the maximum of the 
entropy per particle,
\begin{equation}
\mathcal{S}=-\int d^3r [\rho(r)\ln\rho(r)-\rho(r)]\:+\:\frac{3}{2}
\ln[-\Lambda+\int d^3r d^3r' \rho(r)\rho(r')\bar\phi(r-r')] \, ,
\end{equation}
in the MC, and with the minimum of the free energy per particle,
\begin{equation}
\mathcal{F}=-\int d^3r [\rho(r)\ln\rho(r)-\rho(r)]\:+\:
\eta\int d^3r d^3r' \rho(r)\rho(r')\bar\phi(r-r') \, , 
\end{equation}
in the CE, with the constraint $\int d^3r\rho(r)=1$, if the grid is
fine enough.
The thermodynamical functions depend therefore on two dimensionless
parameters: $s/R$ and $\Lambda$ or $\eta$. However, to derive mean
field theory rigorously, one has to take the limit $N\rightarrow\infty$
in (\ref{dmcpf})~and~(\ref{dcpf}), keeping the number of cells constant. 
Otherwise, if $W$ grows with $N$,
the integrals of Eqs.~(\ref{dmcpf})~and~(\ref{dcpf}) cannot be
evaluated by the saddle point. This implies that $R$ must be kept constant.
Therefore, the factors $N$ entering $\Lambda$ and $\eta$ cannot be 
absorbed in $R$. They can be absorbed in $G$ or in $m^2$.
Hence, mean field theory does not support the diluted limit either.
\\

In conclusion, the thermodynamical limit of a self-gravitating system does
not exist, either in the usual form or in the diluted regime of 
Ref.~\cite{devega}. The meaning of 
\textit{non-existence of the thermodynamical limit} is that 
the thermodynamical potentials do not scale properly with $N$ and
thus thermodynamical functions, such as temperature, diverge.
Nevertheless, it is possible to take the usual thermodynamical limit 
and, consequently,
to use safely the usual thermodynamical tools by first regularizing the
long distance behavior of the gravitational potential, introducing a
very large screening length. The system is then thermodynamically
stable and the thermodynamical limit does exist. Afterwards, one can
study the limit in which the screening length tends to infinity
\cite{victor}.
\\

I thank P.-H. Chavanis and O. Fliegans for some interesting discussions.
This work has been carried out with the financial support of
a Ram\'on y Cajal contract of Ministerio de Ciencia y Tecnolog\'{\i}a
of Spain.


\begin{thebibliography}{}
\bibitem{devega}
H.J. de Vega and N. S\'anchez, Phys. Lett. \textbf{B490}, 180 (2000);
Nucl. Phys. \textbf{B625}, 409 (2002);
\textit{ibid}, 460.
\bibitem{pad}
T. Padmanabhan, Phys. Rep. {\bf 188}, 285 (1990). 
\bibitem{kiess}
M. Kiessling, J. Stat. Phys. \textbf{55}, 203 (1989).
\bibitem{stability}
D. Ruelle, Helv. phys. Acta {\bf 36}, 183 (1963).
M.E. Fisher, Archive for Rational Mechanics and 
Analysis {\bf 17}, 377 (1964). \\
J. van der Linden, Physica {\bf 32}, 642 (1966);
\textit{ibid} {\bf 38}, 173 (1968). 
J. van der Linden and P. Mazur, {\em ibid}
{\bf 36}, 491 (1967).
\bibitem{gravothermal}
V.A. Antonov, Vest. leningr. gos. Univ. {\bf 7}, 135 (1962).
D. Lynden-Bell and R. Wood, Mon. Not. R. Ast. Soc. {\bf 138}, 495 (1968).
W. Thirring, Z. Phys. {\bf 235}, 339 (1970).
\bibitem{chavanis}
P.-H. Chavanis, Phys. Rev. E\textbf{65}, 056123 (2002). P.-H. Chavanis 
and I. Ispolatov, Phys. Rev. E\textbf{66}, 036109 (2002).
\bibitem{edu}
E. Follana and V. Laliena, Phys. Rev. E\textbf{61}, 6270 (2000).
I. Hernquist and N. Katz, ApJS \textbf{70}, 419 (1989).
J. Sommer-Larsen, H. Vedel, and U. Hellsten, ApJ \textbf{500}, 610 (1998).
\bibitem{victor}
V. Laliena, astro-ph/0202448.
\end{thebibliography}
\end{document}